\documentclass[reprint,prb,twocolumn,longbibliography,noeprint,superscriptaddress,floatfix]{revtex4-2}
\usepackage{float}
\usepackage{physics}
\usepackage{graphicx}
\usepackage{xcolor}
\usepackage{amsmath,,amsfonts,amssymb}
\usepackage{textcomp}
\usepackage{bbm}
\usepackage{dsfont}
\usepackage{microtype}
\usepackage{mathtools}
\usepackage{multirow}
   
\usepackage[pdftex,bookmarks=true,bookmarksopen,bookmarksnumbered,
                colorlinks,
                linkcolor=blue,
                citecolor=blue,
                colorlinks = true,
                urlcolor  = blue,
                anchorcolor = blue
                ]{hyperref}
\usepackage{cleveref}

\renewcommand{\figurename}{FIG.}
\makeatletter
\renewcommand*{\fnum@figure}{{\normalfont \figurename~\thefigure}}
\renewcommand*{\@caption@fignum@sep}{ $~$}

\renewcommand{\tablename}{Table}
\makeatletter
\renewcommand*{\fnum@table}{{\normalfont \tablename~\thetable}}

\usepackage{float}
\usepackage{newfloat}
\usepackage{standalone}
\usepackage{booktabs}
\usepackage[geometry]{ifsym}

\crefname{figure}{Fig.}{Figs.}
\crefname{table}{Table}{}
\crefname{section}{Sec.}{Secs.}
\crefname{equation}{Eq.}{Eqs.}

\graphicspath{{paper_figures/}}

\newcommand{\nv}{\texorpdfstring{NV\textsuperscript{-}~}{NV-~}}

\usepackage{lineno}
\begin{document}

\title{High Fidelity Control of a Nitrogen-Vacancy Spin Qubit at Room Temperature using the SMART Protocol}

\author{Hyma H. Vallabhapurapu}
\email[]{h.vallabhapurapu@unsw.edu.au}
\affiliation{
School of Electrical Engineering and Telecommunications, The University of New South Wales, Sydney, NSW 2052, Australia
}

\author{Ingvild Hansen}
\affiliation{
 School of Electrical Engineering and Telecommunications,
 The University of New South Wales, Sydney, NSW 2052, Australia
}

\author{Chris Adambukulam}
\affiliation{
 School of Electrical Engineering and Telecommunications,
 The University of New South Wales, Sydney, NSW 2052, Australia
}

\author{Rainer St\"ohr}
\affiliation{
 Center for Applied Quantum Technology, University of Stuttgart, Stuttgart 70049, Germany
}

\author{Andrej Denisenko}
\affiliation{
Center for Applied Quantum Technology, University of Stuttgart, Stuttgart 70049, Germany
}

\author{Chih Hwan Yang}
\affiliation{
 School of Electrical Engineering and Telecommunications,
 The University of New South Wales, Sydney, NSW 2052, Australia
}

\author{Arne Laucht}
\email[]{a.laucht@unsw.edu.au}
\affiliation{
School of Electrical Engineering and Telecommunications, The University of New South Wales, Sydney, NSW 2052, Australia
}

\begin{abstract}
A practical implementation of a quantum computer requires robust qubits that are protected against their noisy environment. Dynamical decoupling techniques have been successfully used in the past to offer protected high-fidelity gate operations in negatively-charged Nitrogen-Vacancy (NV\textsuperscript{-}) centres in diamond, albeit under specific conditions with the intrinsic nitrogen nuclear spin initialised. In this work, we show how the SMART protocol, an extension of the dressed-qubit concept, can be implemented for continuous protection to offer Clifford gate fidelities compatible with fault-tolerant schemes, whilst prolonging the coherence time of a single NV\textsuperscript{-} qubit at room temperature. We show an improvement in the average Clifford gate fidelity from $0.940\pm0.005$ for the bare qubit to $0.993\pm0.002$ for the SMART qubit, with the nitrogen nuclear spin in a random orientation. We further show a $\gtrsim$ 30 times improvement in the qubit coherence times compared to the bare qubit.
\end{abstract}

\maketitle

\section{\label{sec:introduction} Introduction}

The negatively charged Nitrogen-Vacancy (NV\textsuperscript{-}) centre in diamond \cite{Doherty2013} has long been proposed as a potential qubit platform for room-temperature quantum computing, owing to its long coherence times and compatibility with optically detected magnetic resonance techniques \cite{Wrachtrup2006,Balasubramanian2009,Zhang2018}. The longitudinal relaxation time $T_1$ and transverse relaxation time $T_{2}$, together dictate how long a spin-qubit can remain in a prepared state. These metrics of coherence are of general importance for applications in quantum computing and memory nodes for quantum communication \cite{divincenzo1999quantum,zhong2015optically,bhaskar2020experimental}.

The overall coherence of an \nv{} spin qubit is typically limited by a relatively shorter $T_{2}$ coherence time, due to decoherence produced by a nuclear spin bath \cite{childress} in isotopically unpurified diamond. The nuclear spin bath, along with the host nuclear spin of the \nv{}, therefore negatively impact qubit gate operations, leading to lower gate fidelities \cite{PhysRevLett.112.050502}.

Qubit gate operations with fidelities exceeding 0.99 are said to allow for error-correction strategies which are critical for qubit scalability \cite{dobrovitski2013quantum,knill2005quantum,PhysRevA.86.032324}. To achieve this,  dynamical decoupling techniques may be employed to protect the qubit \cite{de2010universal,PhysRevLett.112.050502,van2012decoherence,PhysRevA.91.052315}. In addition, universal holonomic gates \cite{arroyo2014room,Sekiguchi2022} and composite pulse sequences \cite{rong2015experimental} have also been demonstrated to exhibit high gate fidelities. However, in most cases a static magnetic field (\textbf{B}\textsubscript{0}) aligned with the \nv main symmetry axis is required to polarise the host nitrogen nuclear spin \cite{Chakraborty_2017,Busaite}, which in turn reduces the occurrence of detuned gate operations. While helpful for gate fidelities, initialisation of the nuclear spin presents an added layer of complexity towards scalability, requiring either RF delivery or careful alignment of the \textbf{B}\textsubscript{0} field for multiple \nv spin qubits. If the host nuclear spin is of no interest, it will be useful to perform high-fidelity qubit gate operations on the electronic spin without nuclear spin initialisation.

Alternative control strategies include the implementation of microwave (MW) dressed states, which protect the qubit against magnetic field fluctuations by offering continuous dynamical decoupling \cite{Golter,Morishita2019,Xiangkun,miao2020universal}. The always-on MW field redefines the quantization axis along the driving axis, and the quantization energy according to the driving strength. The latter opens up alternative two-axis control methods \cite{Laucht2016,seedhouse}. Recently, the Sinusoidally Modulated, Always Rotating and Tailored (SMART) protocol as an extension to the dressed protocol, was employed to achieve high fidelity gates in a SiMOS quantum dot device at low temperatures \cite{ingvild1,ingvild2}. In this protocol the resonant MW drive is sinusoidally modulated to reduce the impact of MW detuning and amplitude fluctuations; in particular, the effect of detuned driving has a significant impact on idle qubits (identity gate operation). Some advantages of the SMART protocol over other methods, such as dynamical decoupling sequences, include simplicity in implementation, effective protection against decoherence even during gate operations and compatibility with global control of spin qubits \cite{vahapoglu2021single,vahapoglu2021coherent}. Similar alternating MW fields have also recently been employed to protect qubits in hBN from a nuclear spin bath \cite{ramsay2022room}, and to optically address \nv spins using universal holonomic gates in diamond \cite{Sekiguchi2022} at room temperature.

In this work, we utilize the SMART protocol to demonstrate room temperature, high-fidelity, single qubit gates on the negatively charged Nitrogen-Vacancy (\textsuperscript{15}NV\textsuperscript{-}) electron spin, hosted within an electronic grade CVD diamond sample. We first measure the bare qubit coherence times and compare them to the extended coherence time using the SMART protocol. We then construct various single qubit gates in the bare, dressed, and SMART qubit bases and extract the average Clifford gate fidelities using randomized-benchmarking techniques \cite{muhonen}-- thereby demonstrating high control fidelities compatible with fault-tolerant architectures and the advantage of the SMART protocol.

\section{Qubit Coherence}

\begin{figure*}[ht]
\includegraphics[keepaspectratio]{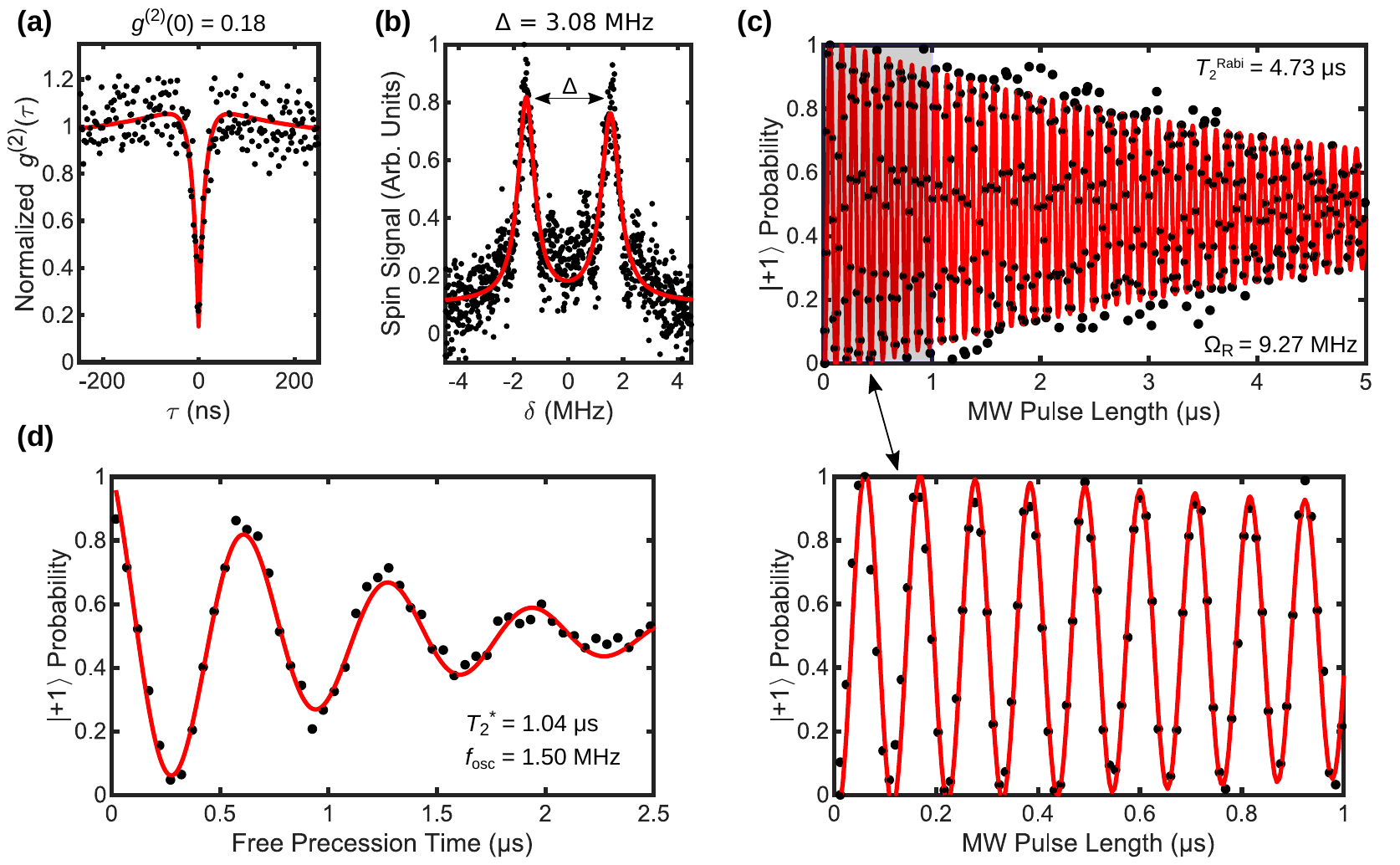}
\caption{Single qubit ODMR. (a) An optical second order correlation measurement showing a normalized $g^{(2)}(0)$ value of 0.18 confirms photon emission from a single \nv centre. (b) Hyperfine levels within the $\ket{0}\leftrightarrow\ket{+1}$ MW transition manifold showing a separation of $\sim$3.1 MHz consistent with literature for \textsuperscript{15}N based \nv defects. (c) Coherent Rabi oscillations observed when driving the $\ket{0}\leftrightarrow\ket{+1}$ electronic transition with the MW frequency equally detuned from both hyperfine levels. (d) Ramsey oscillations observed showing the free-induction decay of the electronic spin at roughly half the hyperfine splitting.}
\label{figure_1}
\end{figure*}

We first implement a bare spin qubit using the $\ket{0}\leftrightarrow\ket{+1}$ spin transition of a single \nv electron spin with a \textsuperscript{15}N isotope nuclear spin ($I=1/2$).  The chemical vapour deposition grown diamond sample used in this work contains etched nanopillars for improved photon collection efficiency and better isolation of single \nv defects with a non-resonant laser \cite{Ali}. The Hamiltonian of the \nv spin is written in the laboratory frame as
\begin{multline}
\label{eq_HS}
    {\cal H_{\rm S}} =D\,\mathbf{S_{\rm z}}^2+\gamma_{\rm e}\,\mathbf{B_0}\cdot\mathbf{S} -\gamma_{\rm n}\,\mathbf{B_0}\cdot\mathbf{I} \\ + A_{\rm ||}\,\mathbf{S_{\rm z}}\,\mathbf{I_{\rm z}}+A_{\rm \perp}(\mathbf{S_{\rm x}}\,\mathbf{I_{\rm x}}+\mathbf{S_{\rm y}}\,\mathbf{I_{\rm y}}),
\end{multline}
where $D$ is the zero-field splitting of $\sim2.87$ GHz, \{\textbf{S},~\textbf{I}\} are operators for the electron and nuclear spins respectively, \textbf{B\textsubscript{0}} is the static magnetic field, \{$\gamma_{\rm e}$, $\gamma_{\rm n}$\} are gyromagnetic ratios for the electron and \textsuperscript{15}N nuclear spins respectively, and \{$A_{\rm \perp}$, $A_{\rm ||}$\} are hyperfine constants. The qubit spin state is correlated with the photoluminescence intensity of the \nv electronic spin, and is measured using suitable optically detected magnetic resonance (ODMR) techniques on a measurement setup similar to previous work \cite{vikas1,yang2021}. A $g^{(2)}(\tau)$ measurement confirms a single \nv spin with $g^{(2)}(0)$ $\ll0.5$ as shown in \cref{figure_1}(a). An ODMR sweep measurement shown in \cref{figure_1}(b) reveals that the detected \nv centre is hyperfine coupled to a \textsuperscript{15}N host nuclear spin with a strength of $\sim$\,3.1 MHz, consistent with literature \cite{Groot_Berning_2021,Dam}.  

\begin{table}[htp]
\caption{Hamiltonian physical constants for \nv{}.}
\begin{center}
\begin{tabular}{cll}
\textbf{Parameter} & \textbf{Value} & \textbf{Description}\\
\hline
\hline
$\gamma_{\rm e}$ & 28 GHz/T & Electron gyromagnetic ratio\\

$\gamma_{\rm n}$ & -4.316 MHz/T & Nuclear gyromagnetic ratio\\

$A_{||} = A_{\perp{}}$ & 3.1 MHz & Isotropic hyperfine\\

$D$ & 2.87 GHz & Zero-field splitting\\

$B_{0}$ & 1.85 mT & Static B field (simulated)\\

\hline
\end{tabular}
\end{center}
\label{ham_constants}
\end{table}%

\subsection{Bare qubit}

A permanent magnet is placed near the diamond sample to Zeeman-split the degenerate $m_{s}=\pm1$ electronic energy levels. Simulation of the \nv Hamiltonian to match the Zeeman-splitting indicates a magnetic field of $\sim$1.8 mT. The hyperfine-coupled host nuclear spin is not initialised in our experiments. To overcome this, we coherently drive the electron spin equally detuned from both hyperfine-split ODMR frequencies with a strong $B_1$ field ($\sim24$ dBm delivered to a thin PCB antenna \cite{vikas1}). We measure Rabi oscillations of the bare qubit with frequency $\Omega_{\rm R}\approx9$ MHz and a coherence time $T_{2}^{\rm Rabi}=4.73$~\textmu s, as shown in \cref{figure_1}(c). Two-axis control of the bare qubit is implemented by making use of IQ modulation of the applied $B_1$ field.

We then measure the bare qubit $T_{2}^{*}$ coherence time using a Ramsey pulse sequence. The resulting data plotted in \cref{figure_1}(d) shows an oscillating signal, fitted to a single decaying sinusoid with frequency $\sim$1.5 MHz approximately equal to the expected detuning, and a transverse free induction decay time of $T_{2}^{*}=1.04$~\textmu s. When compared to the measured $T_{2}^{\rm Rabi}$, this result already hints that the coherence time of a continuously driven spin qubit may be longer than that of an idle spin qubit.

\begin{figure}[ht]
\includegraphics[keepaspectratio]{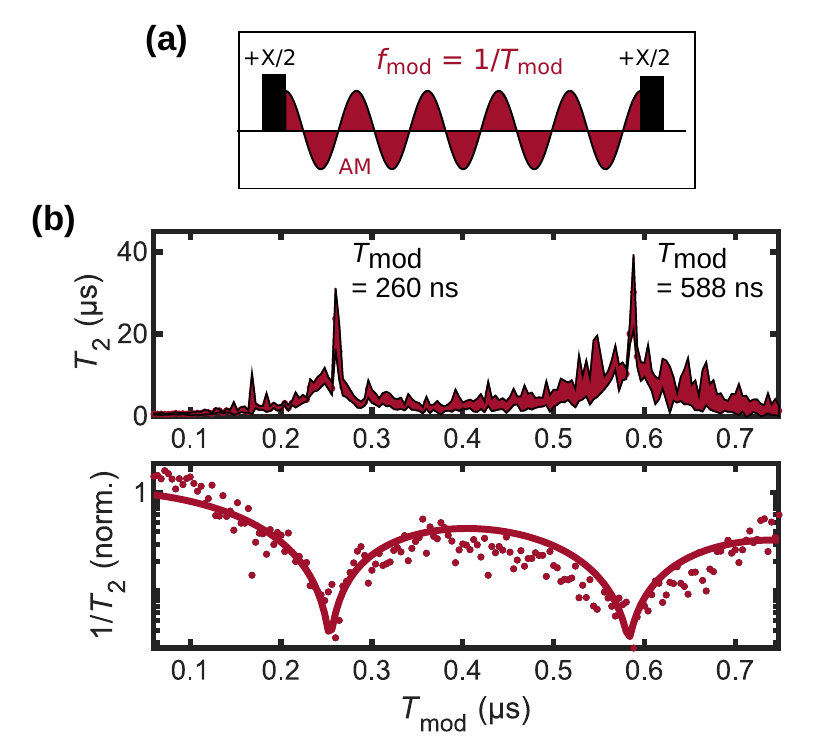}
\caption{A SMART Ramsey measurement. (a) The SMART Ramsey pulse sequence. The $\pm$X/2 pulses and always-on field are applied through the I channel of the MW source. (b) Measured coherence times as a function of modulation period $T_{\rm mod}$. The filled area represents the bounds of the error bars. We find that the coherence time exceeds $\sim30$~\textmu s at optimum modulation periods indicated in the plot. This represents at least a factor 30 improvement in the coherence time compared to the bare spin. The normalised $1/T_{2}$ data is fitted to an absolute valued Bessel function represented by a solid line, showing that the optimum $T_{\rm mod}$ periods are found at the $j_{i}^{\rm th}$ solution of the Bessel function as summarised in \Cref{tab_tmod}.}
\label{figure_2}
\end{figure}

\subsection{Dressed and SMART qubit}

\begin{figure}[htb]
\includegraphics[keepaspectratio]{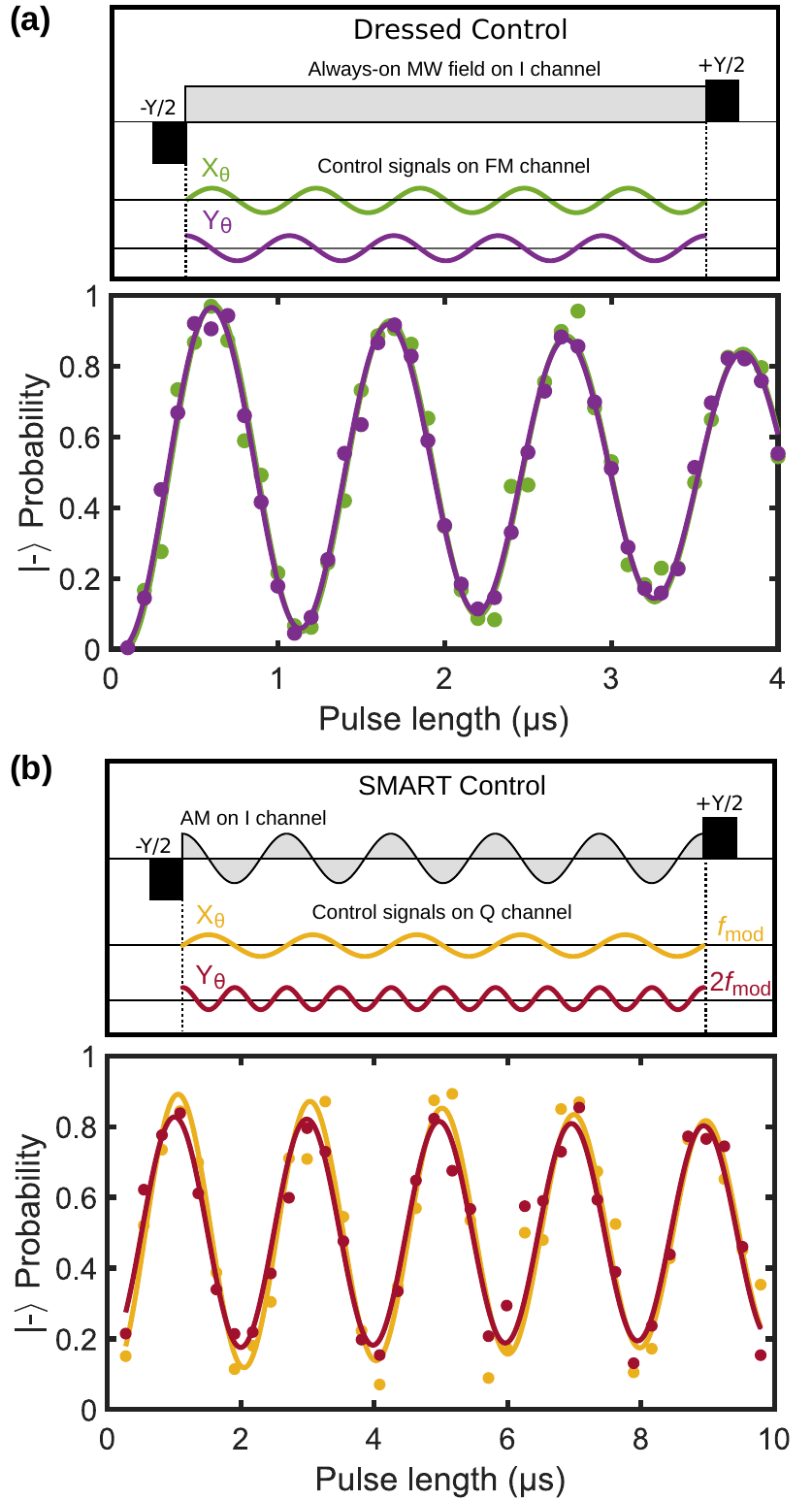}
\caption{Control signals to implement two-axis control for the dressed and SMART qubits. Rabi oscillations indicate coherent $\ket{-}\leftrightarrow\ket{+}$ transitions. (a) We implement frequency modulated control signals, with the FM signal amplitude modulated at a frequency equal to the bare qubit Rabi frequency. The resulting dressed Rabi frequency is then tuned by adjusting the FM depth. (b) To implement two-axis control on the SMART qubit we use amplitude modulation on the IQ MW channels at frequencies $f_{\rm mod}$ and $2f_{\rm mod}$, where the qubit coherence is optimised by using $T_{\rm mod} = T_{opt}$ from \Cref{tab_tmod}.}
\label{figure_3}
\end{figure}

The dressed qubit is implemented by applying an `always-on' resonant MW field to continuously drive the spin; this establishes new eigenstates $\ket{+}$ and $\ket{-}$. It is convenient to transform into the dressed basis (Hadamard transform), in which case the new eigenstates appear along the vertical axis of the Bloch sphere \cite{seedhouse}. The dressed spin is initialised into the $\ket{+}$ state using a -Y/2 gate pulse and projected for measurement using a +Y/2 gate pulse with respect to the bare spin.  

Control of the dressed qubit can then be implemented by detuning pulses via frequency modulation (FM) \cite{arne_dressed_2}, with the X and Y axis control signals being differentiated by a $\pi/2$ phase-shift. The gate lengths are then defined by the FM depth; thus Rabi oscillations in the dressed basis can be observed by sweeping the FM pulse lengths. 

Similarly, to implement the SMART protocol, the always-on MW field is amplitude modulated with a sinusoid where the product of the amplitude and modulation period follows the relationship $\Omega_{\rm R}\times{}T_{\rm mod} = j_{i}$. Here, $\Omega_{\rm R}$ is the bare spin Rabi frequency, $j_i$ is the zeroth order Bessel function and $T_{\rm mod}$ is the amplitude modulation period \cite{ingvild1}.

In order to measure the coherence time of the SMART protocol, we perform a similar Ramsey experiment to that of the bare qubit but replace the free precession with continuous modulated driving. We also sweep the modulation period for increasing driving times (in multiples of $T_{\rm mod}$) while keeping the peak modulation amplitude fixed. The resulting Ramsey signal decay times are fit to the Bessel function, as shown in \cref{figure_2}. The results indicate improved coherence times of $\sim$30~\textmu s and $\sim$40~\textmu s for optimum modulation periods $T_{\rm opt} = 260$~ns and $T_{\rm opt} = 588$~ns, corresponding to $j_{0}$ and $j_{1}$, respectively. The results confirm that the optimal modulation period follows the Bessel function as expected and that the coherence time is improved by at least a factor of 30 compared to the bare qubit. We tabulate the measured and theoretically predicted $T_{\rm mod}$ times in \Cref{tab_tmod}. The small discrepancy between theoretically predicted and measured values for $T_{\rm opt}$ are attributed to slow MW power drifts and detuned Rabi oscillations.

We proceed to implement two-axis control at a sinusoidally modulated MW field with period $T_{\rm mod}=T_{\rm opt}=260$~ns and $\Omega_{\rm R}=9$~MHz. Two-axis control is realised by adding another modulation to an axis perpendicular to the always-on MW driving axis. Hence, with the MW axis applied along X, two-axis control is implemented by adding modulation along Y at frequency $f_{\rm mod}=1/T_{\rm mod}$ (y-gate) and 2$f_{\rm mod}$ (z-gate) \cite{ingvild1}. We therefore apply IQ modulation to implement the SMART qubit control. Two-axis Rabi oscillations of the dressed and SMART qubit are shown in in \Cref{figure_3}(a) and (b), respectively. 

\begin{figure}[htb]
\includegraphics[keepaspectratio]{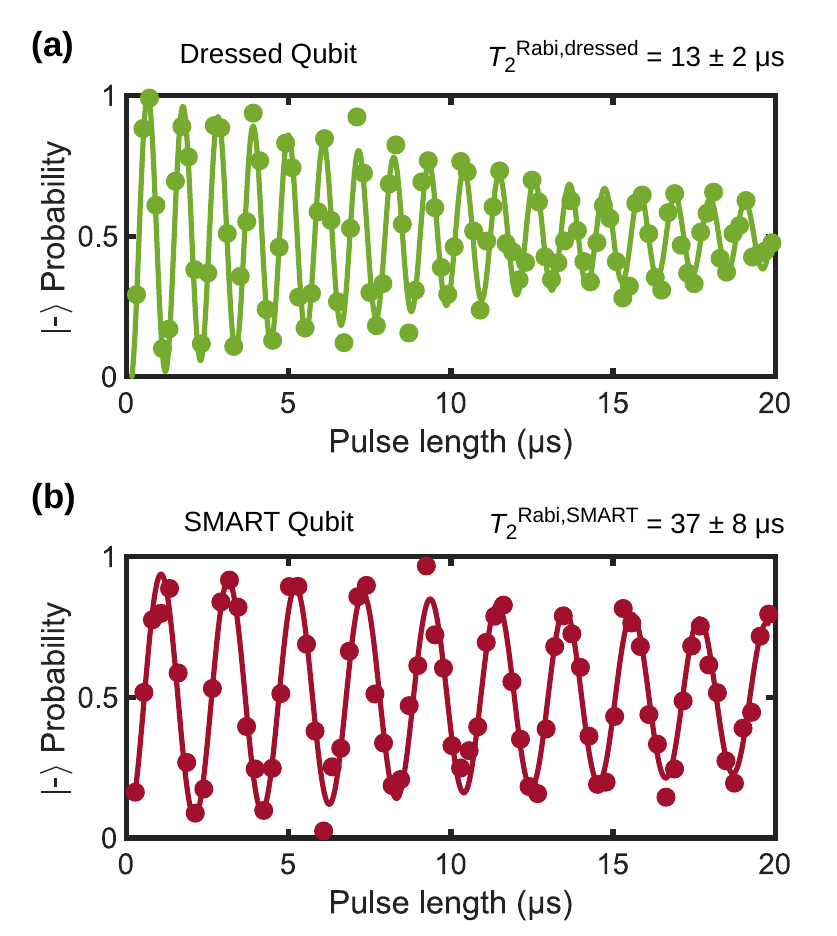}
\caption{Rabi oscillations in the dressed basis similar to \Cref{figure_3} for longer drive times. Compared to the bare spin, the $T_{2}^{\rm Rabi}$ coherence time in the dressed basis may be prolonged by a factor of $\sim3$ using simple qubit dressing, and by a factor of $\sim8$ using the SMART protocol, consistent with coherence improvement shown in \cref{figure_2}. (a) Rabi oscillations of the dressed qubit. (b) Rabi oscillations of the SMART qubit.}
\label{figure_3b}
\end{figure}

In \Cref{figure_3b}, we extend the Rabi experiments to longer pulse lengths and find a significant factor of three improvement in the $T_{2}^{\rm Rabi}$ coherence time of the SMART qubit ($T_{2}^{\rm Rabi,SMART}=37\pm8$~\textmu s) when compared to the dressed qubit ($T_{2}^{\rm Rabi,dressed}=13\pm2$~\textmu s). Such an improvement in coherence time is beneficial for high-fidelity qubit gates, as a higher number of gates may be applied before the qubit experiences appreciable decoherence.

\begin{table}[htb]
\caption{Optimum MW modulation periods $T_{\rm opt}$ for extended qubit coherence.}
\begin{center}
\begin{tabular}{lcccc}
  &  & $\Omega_{\rm R}$ &$T_{\rm opt}$  & $T_{\rm opt}$\\
$i$  &  $j_{i}$ & (MHz) & (theory) & (measured)\\
\hline
\hline
0 & 2.405 &9 & 267 ns & $260$ ns \\
1 & 5.520 &9 & 613 ns & $588$ ns\\
\hline
\end{tabular}
\end{center}
\label{tab_tmod}
\end{table}%

\begin{figure}[htb]
\includegraphics[keepaspectratio]{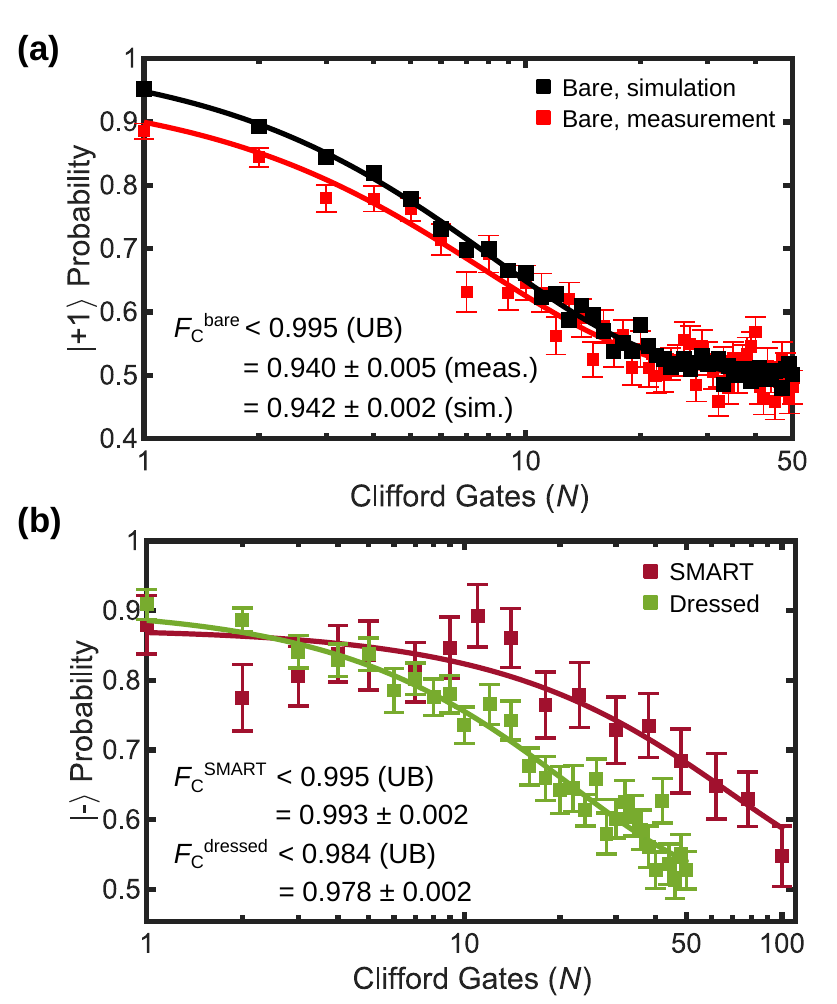}
\caption{Randomized benchmarking of the bare, dressed and SMART qubits. All measurements are based on $\Omega_{\rm R}=9$~MHz on the bare spin. The average Clifford gate fidelity $F_{\rm C}$ is extracted using the relevant fitting function (see text) with 95\% confidence interval. The error bars of individual data points are dominated by signal averaging. Additionally, we calculate the theoretical upper bounds (UB) from the measured Rabi quality factors. 
(a) Measurement of $F_{\rm C}$ for the bare spin qubit is in good agreement with simulation (noise decoherence not included), suggesting that the primary source of low fidelity is the $\sim1.5$ MHz detuning caused by the uninitialised nuclear spin. 
(b) $F_{\rm C}$ for the dressed spin qubit, measured for gates defined by 0.95 MHz dressed Rabi frequency. The fidelity here is expected to be limited by coherence time. 
(c) Similar to the dressed protocol, average Clifford gate fidelities are measured for qubit gates defined by (1/4$T_{\rm mod}$ = 0.96 MHz) SMART Rabi frequencies. We observe $>0.99$ average Clifford gate fidelity for approximately the same Rabi frequency as for the dressed qubit ($\sim$0.95 MHz), demonstrating a significant improvement consistent with enhanced Rabi coherence times.}
\label{figure_4}
\end{figure}

\section{High Fidelity Clifford Gates}

We perform randomized benchmarking on the bare, dressed and SMART qubits to compare their overall average Clifford gate fidelity $F_{\rm C}$, extracted from the fit function $P(N) = P_{0}(2F_{\rm C}-1)^{N}+P_{\infty}$ \cite{muhonen}. Here $N$ is the number of Clifford gates, $P(N)$ is the spin probability as a function of applied Clifford gates, and variables $P_{\infty}$, $P_0$ absorb the state preparation and measurement error~\cite{PhysRevLett.106.180504}. Importantly the variable $P_{\infty}$ is fixed to an experimentally determined value, while  $P_0$ and $N$ are left as free parameters for the fit. 
Additionally, we calculate the theoretical upper bound (UB) for $F_{\rm C}$ using $1-1.875/(4Q_{\rm R}) $\cite{stano2022review,muhonen}, where $Q_{\rm R}$ is the measured Rabi quality factor.

We also simulate randomized benchmarking on the bare qubit using time-domain simulation of the NV\textsuperscript{-} Hamiltonian shown in \Cref{eq_HS}. For the measurements we average over the complete gate space constructed from $\rm \{X,\pm{X/2},Y,\pm{Y/2}\}$ primitive gates for $k\ge30$ repetitions, while for the simulations we average for $k=50$ repetitions to ensure convergence. Additionally, we repeat the measurements for both target output states $\ket{+1}, \ket{0}$ in the bare spin basis and $\ket{+}, \ket{-}$ in the dressed spin basis, and combine the results to yield an average Clifford gate fidelity. The average Clifford gate fidelities for the bare, dressed and SMART qubits are compared in \cref{figure_4}.

As seen in \cref{figure_4}(a), a simulation of the randomized benchmarking for the bare qubit without decoherence modelling(black symbols) is able to accurately predict the measured average Clifford gate fidelity of $F_{\rm C}^{\rm bare}=0.940\pm0.005$ (red symbols). The upper bound for the bare $F_{\rm C}$ calculation is higher than this value ($<$ 0.995) since $Q_{\rm R}^{\rm bare}$ is not affected by detuned driving. These results suggest that the main limiting factor for the bare qubit control fidelity is the detuning of 1.5 MHz due to the uninitialised nuclear spin. The dressed qubit with a 0.95 MHz Rabi frequency however offers a much better gate fidelity of $F_{\rm C}^{\rm dressed}=0.978\pm0.002$; the main limiting factor for the dressed qubit is attributed to spin decoherence. Higher dressed Rabi frequencies may be problematic due to breakdown of the rotating-wave approximation, unless the bare Rabi frequency is also increased~\cite{arne_dressed_2}. For practical difficulties in increasing the MW power, we are limited to the bare Rabi frequency of 9 MHz.

For the SMART qubit, the gate lengths are restricted to a multiple of $T_{\rm opt}$. Thus, with $T_{\rm opt}=260$ ns, we calibrate the SMART Rabi frequency to $1/(4T_{\rm opt})$. The SMART protocol results in even higher average Clifford gate fidelity of $F_{\rm C}^{\rm SMART} = 0.993\pm0.002$, exceeding 0.99, for a similar Rabi frequency of $\sim{}0.96$~MHz, and achieving results similar to \cite{ingvild2}. This is expected as the SMART qubit exhibits an extended coherence time, as can be seen from inspection of \Cref{figure_2} and \Cref{figure_3b}. Although \Cref{figure_2} shows a marginally better coherence improvement for the second $T_{\rm opt}$ operating setpoint of 588 ns, we expect that the reduction in the resulting Rabi frequency would not be of benefit for achieving higher gate fidelities. Future work in this direction may explore using multi-tone amplitude modulation instead to work around this limitation~\cite{ingvild2}.

\section{Conclusion}
High fidelity control of qubits with gate fidelities exceeding 0.99 is critical for implementing various fault-tolerant quantum computing circuits. In this work we have shown that the SMART protocol enables high fidelity control of the electronic spin of an \nv{} spin qubit at room temperature, without needing to initialise the hyperfine-coupled host nuclear spin. We first demonstrate a factor of 30 improvement in the $T_{2}$ coherence time of the SMART qubit when compared to the bare qubit. We then measure an average Clifford gate fidelity of 0.993$\pm$0.002 for the SMART qubit using randomized benchmarking techniques. We find that the SMART protocol achieves gate fidelities comparable to dynamical decoupling methods albeit with a simpler and environment-agnostic implementation, thereby simplifying experimental requirements and qubit operations, and providing a scalable path for multi-qubit operation.

\clearpage
\section*{Acknowledgements}
We acknowledge support from the Australian Research Council (CE170100012). H.H.V. and I.H. acknowledge support from the Sydney Quantum Academy. H.H.V also acknowledges support from the Australian Government Research Training Program Scholarship. C.A. and A.L. acknowledge support from the University of New South Wales Scientia program.

\bibliography{references}
\end{document}